\documentclass[aps,twocolumn,nofootinbib]{revtex4}
\usepackage[latin1]{inputenc}
\usepackage{epsfig}
\newcommand{\beq}{\begin{equation}}
\newcommand{\eeq}{\end{equation}}
\newcommand{\la}{\langle}
\newcommand{\ra}{\rangle}

\begin{document}

\title{Stochastic approach to epidemic spreading}

\author{Tânia Tomé and Mário J. de Oliveira}
\affiliation{Universidade de São Paulo, Instituto de Física,
Rua do Matão, 1371, 05508-090 São Paulo, SP, Brazil}

\begin{abstract}

We analyze four models of epidemic spreading using a
stochastic approach in which the primary stochastic variables
are the numbers of individuals in each class. The stochastic
approach is described by a master equation and the transition
rate for each process such as infection or recovery
are set up by using the law of mass action. We perform
numerical simulations as well as numerical integration
of the evolution equations for the average number of
each class of individuals. 
The onset of the epidemic spreading is obtained by
a linear analysis of the disease free state, from
which follows the initial exponential increase of
the infected and the frequency of new cases. 
The order parameter and the variance in the number of
individuals are also obtained characterizing the onset
of epidemic spreading as a critical phase transition.

\end{abstract}

\maketitle

\section{Introduction}

The theoretical study of the epidemic spreading
\cite{bailey1957,anderson1991,renshaw1991,hastings1997,keeling2008}
started with the employment of ordinary differential equations
of the first order in time, which became known as the deterministic
approach \cite{bailey1957}. 
The individuals of a population are classified in accordance with
their condition in relation to the infectious disease and these
equations give the evolution equations on the number of individuals
belonging in each class. The deterministic approach, however,
do not describe, in an explicit manner, the random fluctuations
occurring in a real epidemic spreading. This observation may
have given way to the need of a stochastic approach to 
the epidemic spreading as that developed by 
Bartlett \cite{bartlett1947,bartlett1949}
and by Bailey \cite{bailey1950,bailey1953}.

A stochastic version of the deterministic model
proposed by Kermack and McKendrick \cite{kermack1927} was developed
by Bartlett in 1949 \cite{bartlett1949}. The model,
called susceptible-infective-removed, 
describes 
the spread of a infectious disease in a community of
individuals who acquire permanent immunization. There are
three classes of individuals: the susceptible, the infective
and the recovered. The approach advanced by Bartlett treated
the numbers of individuals in each class as stochastic
variables from which he developed a time evolution equation
for the generating function corresponding to the probability
distribution of these variables.
 
The evolution equation for the probability distribution,
or master equation, of the model analyzed by Bartlett
was obtained by Bailey \cite{bailey1953}.
The stochastic approach they employed
was based on the use of a continuous time Markov process in
a discrete space in which the variables increase or decrease
by one unit. In 1955, Whittle \cite{whittle1955}
presented a stochastic version of the Kermack and McKendrick
theorem \cite{bailey1957,kermack1927} concerning the outbreak
of an epidemic. According to this theorem, if the density
of the susceptible is smaller than a certain value, the 
the epidemic does not outbreak.

Stochastic versions of deterministic models can be obtained
by transforming the numbers of individuals in each class into
stochastic variables, 
as was the case of the deterministic
susceptible-exposed-infective-removed model proposed by
Dietz \cite{dietz1976} which was transformed into a 
stochastic model allowing its Monte Carlo simulation
\cite{olsen1988}.
One way of achieving the stochastic
versions is to set up a master equation in which case one is
left with the problem of finding the transition rates.
Another way is to add noise in the deterministic equations, 
transforming them into Langevin equations. In this case the
problem is reduced to finding the appropriate type of noise. 
The transition rates and noises, once established,
lead to the several approaches used in the study
of epidemic and population models
\cite{nisbet1982,gabriel1990,grenfell1995,andersoon2000,%
matis2000,allen2015,britton2019}.

The approach we use here to analyze four epidemic models
considers the number of individual in each class
as the primary stochastic variables. It is based on
the use of a master equation and on the law of mass action
to set up the transition rates. This is accomplished by 
using the analogy of the processes in which the individuals
change classes with chemical reactions. After that an
expansion method was used to transform the master equation
into a Fokker-Planck equation \cite{tome2009,tome2015L}. 

More detailed stochastic approaches can be conceived 
if one wishes to take into account the spatial structure
where the individuals live. 
In this case, we may for instance, associate to each individual
a stochastic variable
that takes values corresponding to the condition of an individual
in relation to the disease. This will not be pursued here but
models of this type have
in fact been studied by several authors owing
to their relevance to the spreading of
disease in space and because of their critical
behavior
\cite{harris1974,grassberger1983,ohtsuki1986,satulovsky1994,durrett1995,%
antal2001,dammer2003,souza2010,tome2011,souza2013,tome2015,ruziska2017}.

\section{Evolution equations}

\subsection{Master equation}

The description of the time evolution of a system
by a stochastic approach needs first of all the specification
of the variables that will be used as primary
stochastic variables. A detailed approach 
such as that employed in spatial stochastic model could be used.
Here, we follow a less detailed approach,
which uses as primary stochastic variables
the numbers of individuals belonging in each class.
A class of individuals is its condition with respect to
the infectious disease that we are about to study.
Examples are the classes of susceptible, infected,
removed, and exposed.

To properly set up the stochastic approach, we start by
considering that the individuals of a community interact
with each other in such a way that the epidemic will spread in the
population. One individual does not interact with every
person of the community but interacts with a certain number $N$
of individuals, which is not small but is smaller than the total
number of individuals of the community. 
In accordance with the approach we will use, it suffices
to focus on a neighborhood with $N$ individuals. Its
reciprocal $\varepsilon=1/N$ is understood as a parameter
of the present stochastic approach.

We denote by $n_i$ the number of individuals of the $i$-th 
class within the neighborhood,
and by $n$ the vector whose components are the variables $n_i$.
The vector $n$ is identified as a state of the system.
At each time step of the dynamics, the state $n$
changes to a new value $n'$ and the stochastic
dynamics becomes defined by the transition rates
$W_r(n'|n)$ from state $n$ to state $n'$
corresponding to each process involving the
change of an individual class. The equation that governs
the evolution of the probability distribution $P(n,t)$
of $n$ at time $t$, the master equation, is \cite{tome2015L,kampen1981}
\beq
\frac{d}{dt}P(n) = \sum_r\sum_{n'}\{W_r(n|n')P(n')-W_r(n'|n)P(n)\},
\label{114}
\eeq
where the first summation is over the several processes
and the second summation is over the variables $n_i'$ of
all classes. 

Next we have to set up the transition rates. To this end
we use the analogy of the present problem with that
of chemical kinetics. A class of individuals is
analogous to a chemical species, and a process
of changing class is analogous to a chemical reaction. 
As an example of the analogy, we consider the process
that is always present in the evolution of an infectious
disease. It is the process of infection of a
susceptible (S) individual, who becomes exposed (E),
by an infective (I) individual, represented by
\beq
{\rm S} \stackrel{\rm I}{\longrightarrow} {\rm E},
\label{r3}
\eeq
and understood as the catalytic reaction that transform an S 
into one E by the catalyst I.
In this reaction, the number $n_1$
of the susceptible decreases by one unit, the number of
the infective $n_2$ remains invariant, and the number
of the exposed increases by one unit.
The infection transition rate is
\beq
W_{\rm inf} = b N \left(\frac{n_1}{N}\right)\left(\frac{n_2}{N}\right),
\label{105}
\eeq
where $b$ is the infection rate constant.

If the product of the reaction in (\ref{r3}) is 
the catalyst itself, that is,
\beq
{\rm S} \stackrel{\rm I}{\longrightarrow} {\rm I},
\label{r3a}
\eeq
then the reaction is auto-catalytic,
but the infection rate is still given by equation (\ref{105}).

Another example is the process in which an infective (I)
becomes recovered (R), represented by the spontaneous
reaction
\beq
{\rm I} \longrightarrow {\rm R},
\label{r4}
\eeq
in which the number of infected $n_2$ decreases by one unit
and the number of recovered increases by one unit.
The recovered transition rate is
\beq
W_{\rm rec} = c N \left(\frac{n_2}{N}\right),
\label{106}
\eeq
where $c$ is the recovery rate constant.

The rule that we use to set up a transition rate $W_r$,
which is the reaction rate
corresponding to a certain reaction, is understood as
the application of the law of mass action \cite{tome2015L},
and is given by $W_r=Nw_r$ where
\beq
w_r = k_r q_r,
\eeq
where $k_r$ is the rate constant, and $q_r$ is the product of
the fractions $n_i/N$ of each class of individual appearing as
a reactant, including the catalyst if the reaction is catalytic.

\subsection{Simulation}

Let us discretize the time in intervals equal to $\tau$.
If we denote by $P(n)$ and $P'(n)$ the probability distribution
at time $t$ and $t+\tau$, respectively, then the master equation
can be written in the discretize form as
\beq
P'(n) =  \sum_{n'} T(n'|n) P(n),
\eeq
where
\beq
T(n'|n) = \sum_r p_r q_r(n'|n),
\eeq
$p_r=\tau k_r N$, and the sum of $p_r$ equals one.

The numerical simulation of the master equation is carried
out as follows. At each time step, we choose which
reaction to perform. The reaction is chosen with a probability
$p_r$ which, as we have seen above is proportional to the
corresponding reaction rate constant $k_r$. 
After the reaction has been chosen, it will be in fact
executed with a probability equal to $q_r$. If this is the
case then the numbers $n_i$ will change according to
the chosen reaction. This procedure is repeated a number
of times and a sequence of states is generated, starting
from an initial state.

\subsection{Fokker-Planck equation}

According to the law of mass action the transition rate
$W_r(n'|n)$ associated to a certain reaction is always
written as $W_r=N w_r$ where $w_i$ is a fraction or a product
of fractions $x_i=n_i/N$. In the example given by (\ref{105}),
$w_{\rm inf} = b x_1 x_2$ and in the example given by (\ref{106}),
$w_{\rm rec} = c x_2$. This allows us to write the
the master equation (\ref{114}) in terms of $x$,
\beq
\frac{d}{dt}\rho(x)
= N\sum_r\sum_{x'}\{w_r(x|x')\rho(x') - w_r(x'|x)\rho(x)\}.
\label{48}
\eeq

Usually, the transition rates $W_r(n'|n)$ are such that
the differences $n_i'-n_i$ are small numbers, and in fact,
in the cases that we consider here the differences are
$\pm1$ or zero. This means that the difference $x_i'-x_i$
is of the order $\varepsilon=1/N$, a result that allows us
to expand the quantities on the right-hand side of equation
(\ref{48}), around the state $x$. Performing this expansion
up to second order in $\varepsilon$, the result is the
following Fokker-Planck equation
\beq
\frac{\partial \rho}{\partial t} = 
- \sum_i \frac{\partial f_i\rho}{\partial x_i} 
+ \frac{\varepsilon}2 \sum_{ij}\frac{\partial^2 h_{ij}\rho}
{\partial x_i\partial x_j},
\label{107}
\eeq
where $f_i$ and $h_{ij}$ are functions
of $x$ determined from the transition rates.
The first is related to $w_r$ by
\beq
f_i = \sum_r \nu_{ir}  w_r,
\eeq
where the coefficient $\nu_{ir}$ is the 
variation of $n_i$ in the reaction $r$,
and the second is relate to $w_r$ by
\beq
h_{ij} = \sum_r   \nu_{ir} \nu_{jr}  w_r.
\eeq

We point out that the Fokker-Planck equation (\ref{107})
is equivalent to the set of Langevin equations
\beq
\frac{dx_i}{dt} = f_i + \xi_i,
\label{130}
\eeq
where $\xi_i$ are stochastic variables with the
following properties: $\la \xi(t) \ra=0$ and
\beq
\la \xi_i(t)\xi_j(t')\ra = \varepsilon h_{ij}\delta(t-t').
\eeq
As $h_{ij}$ may depend on $x_i$, the random variables
$\xi_i$ represent a multiplicative noise.

\subsection{Evolution of the averages}

The time evolution of the averages of the various quantities
are obtained from the Fokker-Planck as follows. Let us consider
the average
\beq
\la x_i\ra = \int x_i\rho \,dx.
\eeq
We multiply both sides of the Fokker-Planck equation by $x_i$
and integrate in $x$ to get
\beq
\frac{d}{dt}\la x_i\ra = \la f_i\ra,
\label{110}
\eeq
where we have performed appropriate integration by parts
and considered that $\rho$ vanishes quickly as the limits
of the integral is approached.

Next we determine the time evolution of the covariances
$C_{ij}=\la x_ix_j\ra-\la x_i\ra\la x_j\ra$. To this end we find
first the time evolution of the average $\la x_ix_j\ra$.
We proceed in the same way as above to get the result
\beq
\frac{d}{dt}\la x_ix_j\ra
= \la x_i f_j\ra + \la x_j f_i\ra + \varepsilon \la h_{ij}\ra,
\eeq
from which we find, with the help of (\ref{110})
\beq
\frac{d}{dt} C_{ij} = \la x_i f_j\ra - \la x_i\ra \la f_j\ra
+ \la x_j f_i\ra - \la x_j\ra \la f_i\ra + \varepsilon \la h_{ij}\ra.
\label{108}
\eeq

The equations (\ref{110}) do not consist of a closed set of
equations for the averages $\la x_i\ra$. However,
if $\varepsilon$ is small
we may replace the average $\la f(x)\ra$ by 
$f(\la x\ra)$ on the right-hand side of (\ref{110})
and the set of equations become closed.
The corrections will be of the order $\varepsilon$ and can thus
be neglected. The reasoning to reach this result is as follows.
In the limit $\varepsilon\to0$,
the probability distribution $\rho$ becomes sharped around $x_i$,
giving way to assume that it is a Gaussian distribution
with mean $\la x_i\ra$ and covariances $C_{ij}$, proportional 
to $\varepsilon$. This assumption allows to replace  
$x$ in the average $\la f_i(x)\ra$ by $f(\la x\ra)$
so that equation (\ref{110}) becomes the
equation
\beq
\frac{d}{dt} \bar{x} = f_i(\bar{x}),
\label{112}
\eeq
where we used the simplified notation $\bar{x}_i=\la x_i\ra$.
We see that the evolution equations (\ref{112}) are
now closed equations for the averages $\bar{x}_i$.

Now we use the assumption that the distribution is
a sharped Gaussian distribution to determine the first
terms on the right-hand side of (\ref{108}). But before
we expand 
\beq
\la x_i f_j\ra - \la x_i\ra \la f_j\ra = 
\sum_k f_{jk}(\bar{x})C_{ik},
\eeq 
\beq
\la x_j f_i\ra - \la x_j\ra \la f_i\ra = 
\sum_k f_{ik}(\bar{x})C_{jk},
\eeq 
where $f_{jk}=\partial f_j/\partial x_k$.
Replacing these results in (\ref{108}), we find
\beq
\frac{d}{dt} C_{ij} =
\sum_k \{f_{jk}(\bar{x})C_{ik} + f_{ik}(\bar{x})C_{jk}\}
+ \varepsilon h_{ij}(\bar{x}),
\eeq
which is the equation that determines $C_{ik}$ once
we have determined $\bar{x}_i$, and confirms that
the variances are indeed proportional do $\varepsilon$.
Due to this dependence it is convenient to define
a reduced covariance $\chi_{ij}$ by
$C_{ij}=\varepsilon\chi_{ij}$, which obeys the
equation
\beq
\frac{d}{dt} \chi_{ij} =
\sum_k \{f_{jk}(\bar{x})\chi_{ik} + f_{ik}(\bar{x})\chi_{jk}\}
+ h_{ij}(\bar{x}).
\label{113}
\eeq

The expansion in $\varepsilon$ that we have carried out above
allowed us to find the Fokker-Planck equation (\ref{107})
and its associate Langevin equations (\ref{130}), and 
to reach the equations (\ref{112}) and (\ref{113})
by assuming that the solution of the Fokker Planck 
equation is a Gaussian 
with variances proportional to $\varepsilon$. 
Such an expansion was possible because the transition
rates $w_r$ depend only on the fractions $n_i/N$,
a result that follows from our use of the law of mass action. 
The expansion of the master equation in a
small parameter was developed by van Kampen 
in 1961 by assuming that the solution of the master
equation is a Gaussian with variances proportional
to the expanding parameter \cite{kampen1981,kampen1961,kampen1973}.
It was applied to an epidemic 
model by McNeil \cite{mcneil1972} and also considered
by Nisbet and Gurney \cite{nisbet1982} in population dynamics
under the name of diffusion approximation.

\begin{figure*}
\epsfig{file=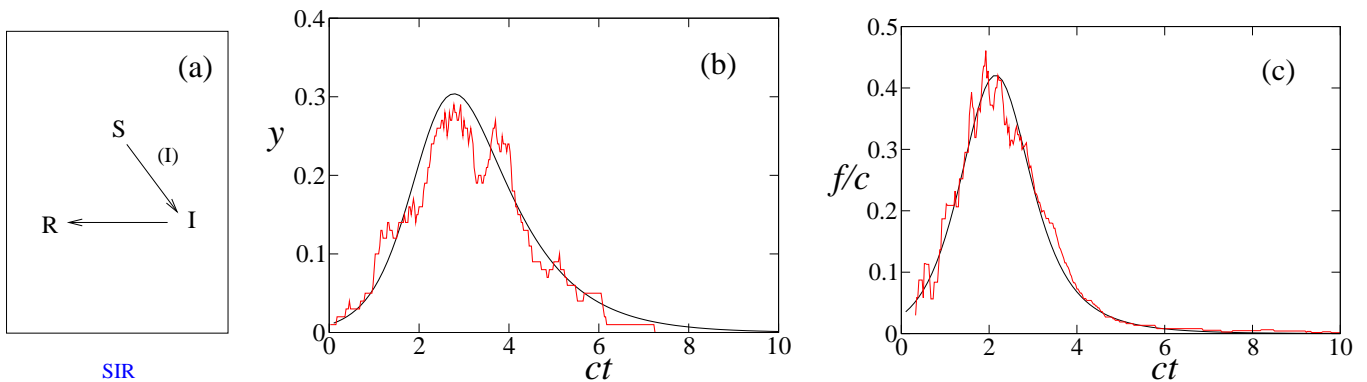,width=3.0cm}
\hfill
\epsfig{file=SIRinf.eps,width=7.0cm}   
\hfill
\epsfig{file=SIRec.eps,width=7.0cm}
\caption{SIR model. (a) The processes composing the model. 
(b) Fraction of infective individuals at a given time
versus time $t$, obtained from numerical simulation
of the master equation and its average.
(c) Epidemic curve from simulation and its average.
All curves were obtained for $b/c=3$.
The simulations were performed using $\varepsilon=0.01$.}
\label{SIR}
\end{figure*}

\section{Critical behavior}

The outbreak of an epidemic is characterized as being a
critical event. If the density of infective individuals
is small there is no spread. But if the density
increases, it will reach a critical density above which
the epidemic spreads, the increase of the infectious 
individuals being exponential in time. This fundamental
idea was used by Ross  in his studies on the transmission
of malaria \cite{ross1911,heesterbeek2015} and was introduced by
Kermack and McKendrick in a clear form as the threshold theorem
\cite{bailey1957,kermack1927}.

To determine the onset of the spread, we perform a
stability analysis of the disease free state, which
is the state without infective individuals. 
This state is always present because the infective
individuals are created catalytically. If the
infective are absent, the system remains forever in
the disease free state, and for this reason it is
called absorbing state in stochastic approaches.

In the present approach, the stability analysis can be
performed by using the evolution equations for the
fractions $\bar{x}_i$ because these equations are
closed equations for these averages. We consider
that the disease free state is a state full of
susceptible individuals so that the fraction 
of the susceptible equals one. The linearization
of the equations (\ref{112}) gives
\beq
\frac{d\bar{x}_i}{dt} = \sum_j f_{ij}\, \bar{x_i},
\eeq
where $f_{ij}=\partial f_i/\partial x_j$ and is calculated at the
disease free state. The equation for the susceptible is
excluded from this set because the equations (\ref{112})
are not in fact all independent as the sum
of the fractions $\bar{x}_i$ equals one.

From the linearized equations it follows that the
time behavior of $\bar{x}_i$ is
\beq
\bar{x}_i = x_{i0}\, e^{\alpha t},
\eeq
where $\alpha$ is the largest eigenvalue of the matrix
with elements $f_{ij}$. The onset of spreading occurs
when $\alpha=0$. When $\alpha>0$ the increase in $\bar{x}_i$
is exponential.

The largest eigenvalue $\alpha$ of the stability matrix
has a relationship with the reproduction number, used to
characterize the epidemic spreading. This quantity
is related to the number of individuals that
can be infected by one individual in a population of susceptible.
It is defined more precisely as follows. Let $N_{\rm a}$
be the number of new cases occurring in a time interval $\Delta t$,
which is given by $N_{\rm a}=N \bar{f} \Delta t$, where $f$ is the
frequency of new cases, that is, $f$ is the fraction of 
individuals that are being infected per unit time. 

The frequency of new cases comes from all reactions of the type
\beq
{\rm A} \stackrel{\rm I}{\longrightarrow} {\rm B},
\eeq
where A represents an individual free of disease and B
an individual that has been infected. Since this reaction
is catalytic and the infective is the catalyst, the
reaction rate is proportional to the fraction of
the infective. Therefore, $f$ is proportional to the fraction
of the infective $y$, that is, $f=g y$, where $g$ depends
on the fractions of the other classes but not on $y$.

Next, we have to determine the number $N_{\rm b}$ of infective
individuals that have infected the $N_{\rm a}$ individuals in
the interval $\Delta t$. If the number of infective remain the
same in the interval $\Delta t$, then $N_{\rm b}$ would be
equal to $N_{\rm a}$. However, the number of infective may
have increased by an amount $N_{\rm c}$ in the interval $\Delta t$,
in which case $N_{\rm b}=N_{\rm a}-N_{\rm c}$. As 
$N_{\rm c}=N(d\bar{y}/dt)\Delta t$ and
\beq
N_{\rm b} = N \bar{f} \Delta t,
\eeq
we get
\beq
N_{\rm b} = N \left(\bar{f} - N \frac{d\bar{y}}{dt}\right) \Delta t,
\eeq
and the reproduction number $R=N_{\rm a}/N_{\rm b}$ becomes
\beq
R = \frac{\bar{f}}{\bar{f}-d\bar{y}/dt}.
\eeq

In the early stages of the epidemic, the reproduction number
is called basic reproduction number, denoted $R_0$. 
In this case the fraction of infective behaves exponentially
with time, $\bar{y}=y_0 e^{\alpha t}$ and 
\beq
R_0 = \frac{\bar{f}}{\bar{f}-\alpha \bar{y}}
= \frac{\bar{g}}{\bar{g}-\alpha},
\label{124}
\eeq
where the second equality was obtained by recalling that
$\bar{f}=\bar{g}\bar{y}$, and $\bar{g}$ is calculated using
the disease free solution.
The onset of spreading occurs when $\alpha=0$, that is, when
$R_0=1$. When $\alpha>0$, that is, when $R_0>1$, the epidemic
spreads whereas when $\alpha<0$, that is, when $R_0<1$, it
does not. The epidemic spreading
occurs when the basic reproduction number is larger than one.

\section{SIR model}

The susceptible-infective-recovered (SIR) model consists of
three classes of individuals, susceptible, infective,
and recovered. The recovered individuals acquires permanent
immunization and cannot be infected again. The model has two
processes. The first is the infection of a susceptible
by an infective, represented by the auto-catalytic reaction
\beq
{\rm S} \stackrel{\rm I}{\longrightarrow} {\rm I},
\label{r7}
\eeq
occurring with an infection rate constant $b$,
and the second is the spontaneous recovery of an infective,
represented by
\beq
{\rm I} \longrightarrow {\rm R},
\eeq
occurring with a recovery rate constant $c$.
In figure \ref{SIR}a we show a representation of the model
involving these two processes.

We denote by $x$, $y$ and $z$ the fractions of the susceptible,
the infected and the recovered, respectively. 
The rate of the infection process is
\beq
w_{\rm inf} = b x y,
\eeq 
whereas the rate of the recovery process is
\beq
w_{\rm rec} = c y.
\eeq
According to the rules above the equations that give
the time evolution of the averages $\bar{x}$, $\bar{y}$,
and $\bar{z}$ are
\beq
\frac{d\bar{x}}{dt} = - b\bar{x}\bar{y}, 
\label{115a}
\eeq
\beq
\frac{d\bar{y}}{dt} = b\bar{x}\bar{y} - c \bar{y},
\label{115b}
\eeq
\beq
\frac{d\bar{z}}{dt} = c \bar{y}.
\label{115c}
\eeq
We remark that these three equations are not independent
because $\bar{x}+\bar{y}+\bar{z}=1$.

We have solved numerically this set of equation and obtained
$\bar{x}$, $\bar{y}$, and $\bar{z}$ as functions of $t$.
In figure \ref{SIR}b we show $\bar{y}$ as a function of $t$
together with $y$ obtained from a simulation of the 
master equation obtained with $\varepsilon=0.01$. The infective
increases exponentially, reaches a maximum and then
decreases towards zero.

The fraction of individuals that are being 
infected per unit time $f$, or frequency of new cases,
is obtained from the infection process (\ref{r7}) and is given by $f=bxy$.
From the simulation, we have obtained
$f$ which is shown in figure \ref{SIR}c together with
its average $\bar{f}=b\bar{x}\bar{y}$ as a function of
time, the epidemic curve. The frequency
of new cases increases exponentially, reaches a maximum
and then decreases towards zero, indicating that the
disease became extinct. 

The initial exponential increase in the fraction of infected,
and thus in the frequency of new cases, is shown by
a stability analysis of the disease free state.
This state corresponds to the absence of disease, 
and all individuals are susceptible. That is, $\bar{x}=1$, 
$\bar{y}=0$ and $\bar{z}=0$, which is a stationary solution
of the set of equations above. As only two equations are
independent, we will use only the last two, which after
linearization gives
\beq
\frac{d\bar{y}}{dt} = \alpha \bar{y}, 
\eeq
\beq
\frac{d\bar{z}}{dt} = c \bar{y},
\eeq
where $\alpha= b-c$.
The solution of the first equation gives
\beq
\bar{y} = y_0 e^{\alpha t},
\eeq
and we see that if $\alpha>0$ then $\bar{y}$ increases exponentially.
The value $\alpha=0$, that is, $b=c$ gives the onset of the
spread because if $\alpha<0$ then $\bar{y}$ dies out.

\begin{figure}
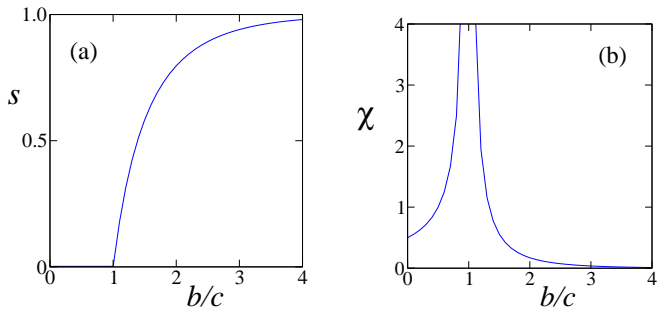

\epsfig{file=SIRop.eps,width=4.0cm}
\hfill
\epsfig{file=SIRchi.eps,width=4.0cm}
\caption{SIR model. (a) Order parameter $s$, which is the area
under the epidemic curve, as a function of $b/c$.
(b) Variance $\chi$ related to the susceptible as a function of $b/c$.}
\label{SIRop}
\end{figure}

As one increases the infection rate constant $b$, from a small
value, it will reach a critical value $b_c=c$ at which the 
spread occurs. The order parameter $s$ of the epidemic spreading
phase is the area under the epidemic curve, that is,
\beq
s = \int_0^\infty \bar{f} dt.
\eeq
In the present case, $\bar{f}=b\bar{x}\bar{y}$ and
from equation (\ref{115a}) we see that $\bar{f}=-d\bar{x}/dt$
and we may conclude that 
\beq
s = 1-x^*,
\eeq
where $x^*$ denotes the value of $\bar{x}$ for long times and
we have taken into account that at initial times $\bar{x}$
equals one.

The basic reproduction number is obtained from
(\ref{124}) and considering that $\bar{f}=b\bar{x}\bar{y}$, we find
\beq
R_0 = \frac{b}{b-\alpha} = \frac{b}{c}, 
\eeq
where we have taken into account that for the disease free
state $\bar{x}=1$ and in the second equality we have used
the result $\alpha=b-c$.

If we divide equations (\ref{115c}) and (\ref{115a}), we find
\beq
\frac{d\bar{z}}{d\bar{x}} = - \frac{c}{b\bar{x}},
\eeq
which after integrating gives
\beq
\bar{z} = -\frac{c}{b}\ln\bar{x},
\label{137}
\eeq
where the integration constant was found by using the disease free
state $\bar{x}=1$ and $\bar{z}=0$. If we denote by $x^*$, $y^*$, and
$z^*$ the values of the fractions for large times,
we see that $x^*+z^*=1$ because $y^*=0$. Therefore an equation
for $z^*$ is obtained by replacing
$z$ by $z^*$ and $x$ by $1-z^*$ in the equation (\ref{137}).
But $s$ equals $1-x^*=z^*$ as we have seen above, so that
\beq
s = -\frac{c}{b}\ln(1-s).
\eeq
This equation gives the order parameter as a function of $b$
and is shown if figure \ref{SIRop}a.
If $b\leq c$, $s$ vanishes. For $b>c$, $s$ is nonzero and 
for $b$ near its critical value $b_c=c$, it is given by
\beq
s = \frac2c(b-c).
\eeq
The order parameter $s$ increases monotonically with infection
rate $b$ from its zero value at the critical point $b_c=c$,
approaching the asymptotic value $z=1$. 

\begin{figure*}
\epsfig{file=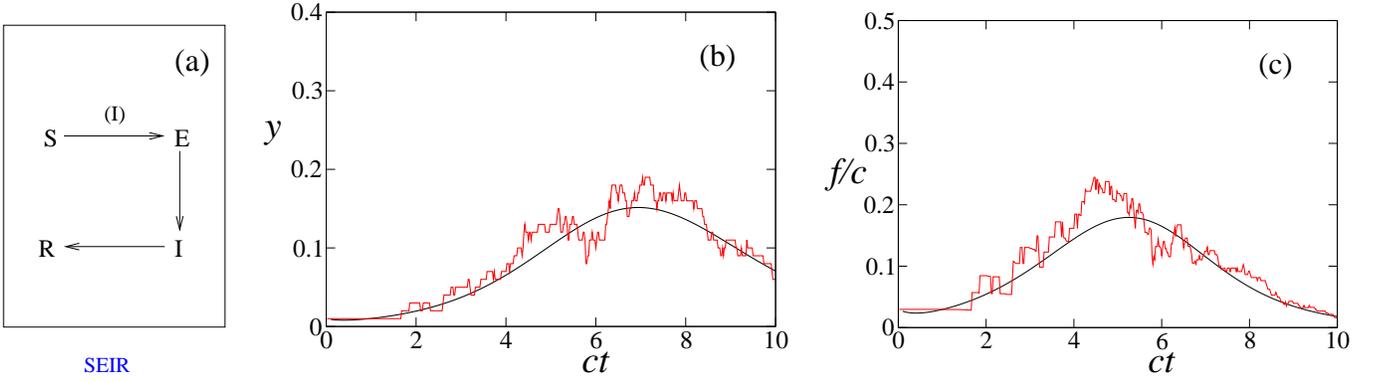,width=3.0cm}
\hfill
\epsfig{file=SEIRinf.eps,width=7.0cm}   
\hfill
\epsfig{file=SEIRec.eps,width=7.0cm}
\caption{SEIR model. (a) The processes composing the model. 
(b) Fraction of infective individuals at a given time
versus time $t$, obtained from numerical simulation
of the master equation and its average. The fraction of the
infective vanishes for long times.
(c) Epidemic curve, or frequency of new cases as a function of time,
from simulation, and its average.
All curves were obtained for $b/c=3$ and latent period $\ell=c/k=1$.
The values of $b/c$ is the same as that used in figure \ref{SIR}
but due to a nonzero value of the latent period the curves 
present a flattening when compared to those of figure \ref{SIR}.
The simulations were performed using $\varepsilon=0.01$.}
\label{SEIR}
\end{figure*}

The use of a stochastic approach allows us to determine
the fluctuations in the variables $x$, $y$, and $z$.
A measure of the fluctuations are given by the covariances.
Using the formula (\ref{113}) for the reduced covariances, 
we find the following expression for the 
reduced variance $\chi$ of the fraction of the susceptible
at the stationary state
\beq
\chi = \frac{c(1-s)}{2(bs-b+c)}.
\eeq
A plot of $\chi$ versus $b$ is shown in figure \ref{SIRop}b.
Near the critical point, it diverges as
\beq
\chi = \frac{c}{2|b-c|}.
\eeq

\section{SEIR model}

There are some diseases such that the suscpetible individuals
that have been infected takes a certain time to be infective.
These individuals, that have got the disease but are not 
capable of infect others, are called exposed. 
The model susceptible-exposed-infective-re\-cov\-ered (SEIR)
is similar to the SIR model but there is an
intermediate step before a susceptible becomes infective
as shown in figure \ref{SEIR}a. The process of infection
is represented by 
\beq
{\rm S} \stackrel{\rm I}{\longrightarrow} {\rm E},
\label{r33}
\eeq
occurring with an infection rate constant $b$,
the process of becoming infective is represented by
\beq
{\rm E} \longrightarrow {\rm I},
\eeq
occurring with a rate constant $k$, and the process
of recovering is represented by
\beq
{\rm I} \longrightarrow {\rm R},
\eeq
occurring with a recovering rate constant $c$.
The inverse of the rate constant $k$ is a measure of the
latent period $\ell$ of the exposed individual.
When the latent period vanishes, $\ell=0$, the present
model reduces to the SIR model, in which a susceptible
that has been infected becomes infective immediately.

We use the same notation as that of the SIR model,
namely, $x$, $y$, and $z$ for the fraction of
susceptible, infective and recovered, and $u$ for the
fraction of the exposed. The rate of the infection process is
\beq
w_{\rm inf} = b x y,
\eeq 
the rate of the process becoming infective is
\beq
w_{\rm ive} = k u,
\eeq 
and the rate of the recovery process is
\beq
w_{\rm rec} = c y.
\eeq
According to the rules, the evolution equation for the
averages of these quantities are
\beq
\frac{d\bar{x}}{dt} = - b\bar{x}\bar{y},
\label{120a}
\eeq
\beq
\frac{d\bar{u}}{dt} = b\bar{x}\bar{y} - k \bar{u},
\label{120b}
\eeq
\beq
\frac{d\bar{y}}{dt} = k \bar{u} - c \bar{y},
\label{120c} 
\eeq
\beq
\frac{d\bar{z}}{dt} = c\bar{y}.
\label{120d}
\eeq
These equations are not all independent because 
$\bar{x}+\bar{u}+\bar{y}+\bar{z}=1$.

We have solved numerically this set of equation and obtained
$\bar{x}$, $\bar{y}$, $\bar{z}$, and $\bar{u}$.
In figure \ref{SEIR}b
we show $\bar{y}$ as a function of time together
with $y$ obtained from the simulation of the master equation
with $\varepsilon=0.01$. The infective growth exponentially, attain a
maximum and then decrease towards the zero value.
In figure \ref{SEIR}c we show the frequency of new cases $f$
which comes from the infective process (\ref{r33}) and is given $f = bxy$.
Its average is $\bar{f}=b\bar{x}\bar{y}$ and is also shown in
the same figure.

\begin{figure*}
\epsfig{file=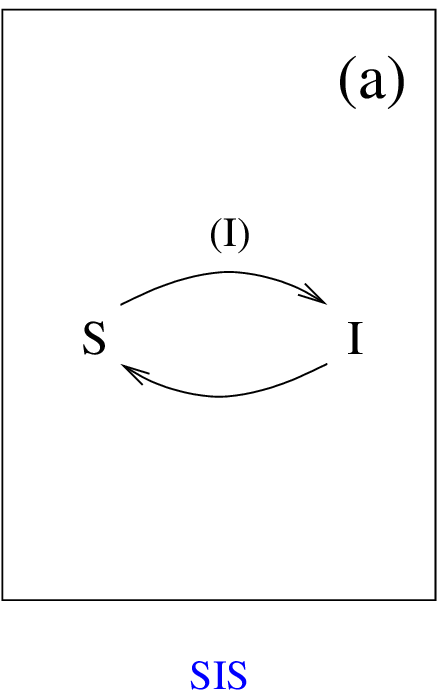,width=3.0cm}
\hfill
\epsfig{file=SISinf.eps,width=7.0cm}   
\hfill
\epsfig{file=SISec.eps,width=7.0cm}
\caption{SIS model. (a) The processes composing the model.
(b) Fraction of infective individuals at a given time
versus time $t$, obtained from numerical simulation
of the master equation and its average. The fraction of the
infective approaches a nonzero asymptotic value.
(c) Epidemic curve, or frequency of new cases as a funtion of time,
from simulation, and its average. For long times it approaches a
nonzero value.
All curves were obtained for $b/c=2$.
The simulations were performed using $\varepsilon=0.01$.}
\label{SIS}
\end{figure*}

We determine now the conditions for the outbreak of the epidemic.
To this end we employ a stability analysis of the disease free
state, which is $\bar{x}=1$, $\bar{y}=0$, $\bar{z}=0$, and
$\bar{u}=0$. After linearizing the equations (\ref{120b}) and
(\ref{120c}) become
\beq
\frac{d\bar{u}}{dt} = b\bar{y} - k \bar{u},
\eeq
\beq
\frac{d\bar{y}}{dt} = k \bar{u} - c \bar{y}.
\eeq
Assuming solutions of the type $\bar{y}=y_0e^{\alpha t}$ and
$\bar{u}=u_0e^{\alpha t}$, we find
\beq
- k u_0 + by_0  = \alpha u_0,
\eeq
\beq
k u_0 - c y_0 = \alpha y_0,
\eeq
which is a set of eigenvalues equations. The largest eigeinvalue is
\beq
\alpha = \frac12\{ -(k+c) + \sqrt{(k-c)^2+4bk} \}.
\label{139}
\eeq
The epidemic spreads when $\alpha>0$ which occurs when $b>c$, and
the threshold of spread occurs when $\alpha=0$, that is, when $b=c$,
results that are independent of $k$.
We see that the process S$\to$E, which occurs with a rate constant $k$,
does not change the outbreak of the epidemic but yields a flattening of
the epidemic curve as seen in figure \ref{SEIR}.

Although the presence of a latent period induces a flattening of the
epidemic curve its area does not change and is the same as that
of the SIR model. To show this result we recall that the frequency
of new cases is $\bar{f} = b\bar{x}\bar{y}$ and from equation (\ref{120a})
we see that $\bar{f} = -d\bar{x}/dt$. Therefore
\beq
s = \int_0^\infty \bar{f} dt = 1- x^*.
\eeq
Now we have to show that $x^*$ does not depend on $k$.
Dividing equations (\ref{120d}) and (\ref{120a}),
\beq
\frac{d\bar{z}}{d\bar{x}} = - \frac{c}{b\bar{x}},
\eeq
which after integration gives
\beq
\bar{z} = - \frac{c}{b}\ln\bar{x},
\label{123}
\eeq
and we recall that $\bar{x}+\bar{y}+\bar{z}+\bar{u}=1$.
For large times the infective as as well as the exposed
disappers, $y^*=0$ and $u^*=0$ and $x^*=1-z^*$. Replacing 
this last result in (\ref{123}) we get an equation for
$x^*$ that does not depend on $k$. The equation for
$s=1-x^*$ follow immediately and is 
\beq
s = - \frac{c}{b}\ln(1-s),
\eeq
and does not depend on $k$ and is the same as that of the SIR model.

As we have seen above the frequency of new cases comes from the
infection reaction (\ref{r33}) as is given by $f=bxy$.
The basic reproduction number is obtained from
(\ref{124}) and given by
\beq
R_0 = \frac{b}{b-\alpha}.
\eeq
Replacing $\alpha$ given by (\ref{139}), we obtain $R_0$
in terms of the rate constant $b$, $c$ and $k$.
The onset of the epidemic spreading occurs when $\alpha=0$,
that is, when $R_0=1$. When $\alpha>0$, the value of $R_0$
is greater than one. It should be remarked that $R_0$
is smaller that the basic reproduction number for the
SIR model. To reach this result, it suffices to recall
that $\alpha_{\rm sir}=b-c$ and that we can show 
from the expression (\ref{139}) that $\alpha\leq b-c$ if $b\geq c$. 
The depression on the basic reproduction number
is a consequence of the time it takes for the exposed
to become infective.

\section{SIS model}

In the two models that we have analyzed above, the infective as well
as the frequency of new cases vanish in the long term. The disease
becomes extinct within the population. In the
susceptible-infective-susceptible model (SIS), the disease does not
disappear, becoming endemic. For long times the infective does
not disappear and the frequency of new cases is nonzero.
The SIS model has only two classes, the susceptible and
the infective, and two processes, as shown in figure \ref{SIS}a.
The first is the infection process represented by 
\beq
{\rm S} \stackrel{\rm I}{\longrightarrow} {\rm I},
\label{r37}
\eeq
occurring with an infection rate constant $b$,
and the recovering process
\beq
{\rm I} \longrightarrow {\rm S},
\eeq
occurring with a recovery rate constant $c$.

We denote by $x$ and $y$ the fraction of
susceptible and infective, respectively.
The rate of the infection process is
\beq
w_{\rm inf} = b x y,
\eeq 
and the rate of the recovery process is
\beq
w_{\rm rec} = c y.
\eeq
According to the rules, the evolution equation for the
averages of these quantities are
\beq
\frac{d\bar{x}}{dt} = - b\bar{x}\bar{y} + c \bar{y},
\eeq
\beq
\frac{d\bar{y}}{dt} = b\bar{x}\bar{y} - c \bar{y}.
\eeq

These equations are not all independent because 
$\bar{x}+\bar{y}=1$. It is convenient to replace $\bar{x}$
in the second equation to obtain just one equation in $\bar{y}$,
\beq
\frac{d\bar{y}}{dt} = \alpha \bar{y}- b\bar{y}^2,
\label{128}
\eeq
where $\alpha=b-c$.

The solution for $\bar{y}$ can be given in closed form,
\beq
\bar{y} = \frac{\alpha y_0}{b y_0 + (\alpha-by_0) e^{-\alpha t}}.
\label{118}
\eeq
In figure \ref{SIS}b we show $\bar{y}$ as a function of $t$
together with $y$ obtained from simulations of the master
equation using $\varepsilon=0.01$. We see that the fraction of
infective does not decrease, remaining finite at large times.
We also show in figure \ref{SIS}c the frequency of new
cases $f=bxy=b(1-y)y$ obtained from simulations as well as
its average $\bar{f}=b(1-\bar{y})\bar{y}$ where $\bar{y}$
is given by the solution (\ref{118}). The frequency of new
cases does not decrease for long times and remains finite.

\begin{figure}
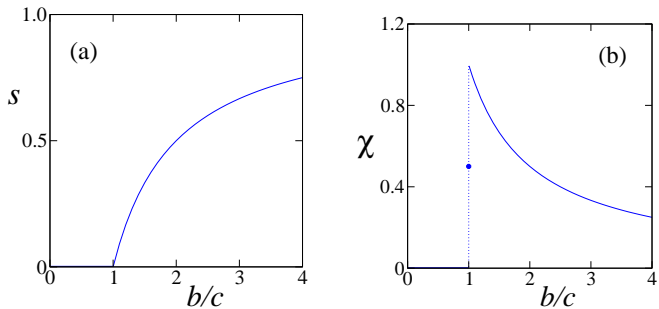

\epsfig{file=SISop.eps,width=4.0cm}
\hfill
\epsfig{file=SISchi.eps,width=4.0cm}
\caption{SIS model. (a) Order parameter $s$, which is the
final value of the fraction of the infective, as a function of $b/c$.
(b) Variance $\chi$ related to the infective a a function of $b/c$.}
\label{SISop}
\end{figure}

The linear exponential increase of $\bar{y}$
can be perceived from the closed solution.
Alternatively, we may obtain this behavior by 
the linearization the equation (\ref{128})
around the disease free solution $\bar{y}=0$, 
\beq
\frac{d\bar{y}}{dt} = \alpha \bar{y},
\eeq
from which follows the solution
\beq
\bar{y} = y_0 e^{\alpha t}.
\eeq
Thus if $\alpha>0$, that is, if $b>c$, the epidemic spreads,
otherwise it does not.
If one increases $b$ from small values, it will reach a 
critical value $b_c=c$ which determined the onset of spread.
The basic reproduction number is obtained from
(\ref{124}) and considering that $\bar{f}=b\bar{x}\bar{y}$,
and that $\alpha=b-c$, we find
\beq
R_0 = \frac{b}{c}.
\eeq

In the limit $t\to\infty$, $\bar{y}$ does not vanish but reaches the
value
\beq
y^* = \frac{b-c}{b}.
\eeq
This value is obtained either by taking the limit $t\to\infty$
in equation (\ref{118}) or by setting to zero the right-hand side
of (\ref{128}), and is identified as the order parameter $s$.
Therefore $s$ is given by
\beq
s = \frac{b-c}{b},
\eeq
and is shown in figure \ref{SISop}a as a function of
the infection rate $b$.

Applying the formula (\ref{113}) for the
present case, we find the following expression for the 
reduced variance $\chi$ of the fraction of the infective
at the stationary state
\beq
\chi = \frac{c}{b}, \qquad b>c,
\eeq
$\chi=0$ for $b<c$, and $\chi=1/2$ when $b=c$.
A plot of $\chi$ versus $b$ is shown in figure \ref{SISop}b.

\begin{figure*}
\epsfig{file=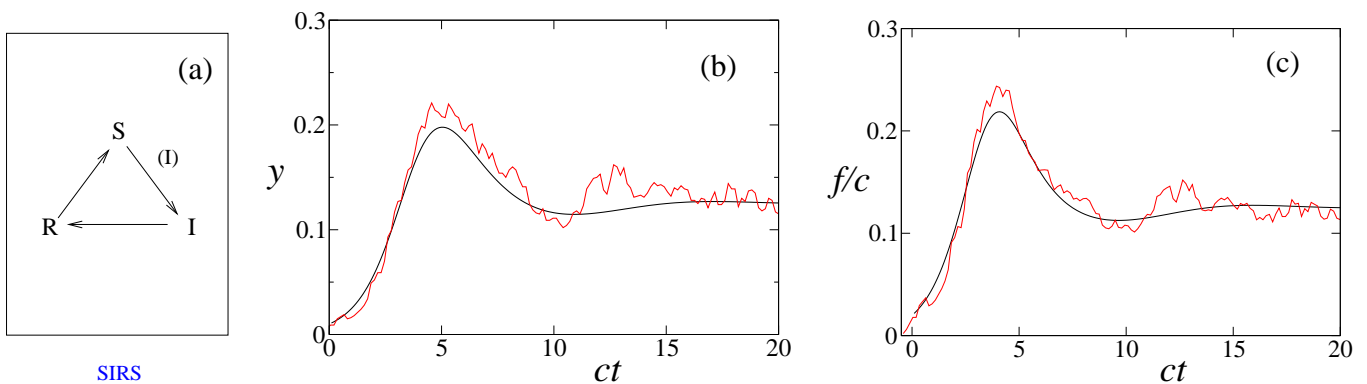,width=3.0cm}
\hfill
\epsfig{file=SIRSinf.eps,width=7.0cm}   
\hfill
\epsfig{file=SIRSec.eps,width=7.0cm}
\caption{SIRS model. (a) The processes composing the model.
(b) Fraction of infective individuals at a given time
versus time $t$, obtained from numerical simulation
of the master equation and its average. The fraction of the
infective approaches a nonzero asymptotic value.
(c) Epidemic curve, or frequency of new cases as a funtion of time,
from simulation, and its average. For long times it approaches a
nonzero value.
All curves were obtained for $b/c=2$ and $a/c=1/3$.
The simulations were performed using $\varepsilon=0.001$.}
\label{SIRS}
\end{figure*}

\section{SIRS model}

In the model we consider now, the infective and the frequency
of new cases do not vanish in the long term and in this sense
it is similar to the SIS model. The
susceptible-infective-recovered-susceptible (SIRS) model
has three classes of individuals
like the SIR mode, susceptible, infective, and recovered,
and one more process than the SIR model. The processes are
are shown in figure \ref{SIRS}a and is as follows.
The infection of a susceptible individual,
\beq
{\rm S} \stackrel{\rm I}{\longrightarrow} {\rm I},
\label{r25}
\eeq
occurring with a rate constant $b$, the spontaneous
recovery,
\beq
{\rm I} \longrightarrow {\rm R},
\eeq
occurring with a rate constant $c$, and
the spontaneous loss of immunity,
\beq
{\rm R} \longrightarrow {\rm S},
\eeq
occurring with a rate constant $a$. 
The recovered individual have only partial immunity
in contrast to the SIR model where the recovered
individual has permanent immunity.

The fractions of susceptible, infective and recoverd
are denoted by $x$, $y$, and $z$, respectively.
The rate of the infection process is
\beq
w_{\rm inf} = b x y,
\eeq 
and the rate of the recovery process is
\beq
w_{\rm rec} = c y,
\eeq
and the rate of the loss of immunity is
\beq
w_{\rm los} = az.
\eeq
According to the rules, the evolution equation for the
averages of these quantities are
\beq
\frac{d\bar{x}}{dt} = - b \bar{x}\bar{y} + a \bar{z},
\label{125a}
\eeq
\beq
\frac{d\bar{y}}{dt} = b\bar{x}\bar{y} - c\bar{y},
\label{125b}
\eeq 
\beq
\frac{d\bar{z}}{dt} = c\bar{y} - a \bar{z},
\label{125c}
\eeq
and they are not all independent because
$\bar{x}+\bar{y}+\bar{z}=1$. 

We have solved numerically this set of equations
and obtained $\bar{x}$, $\bar{y}$, and $\bar{z}$.
In figure \ref{SIRS}b we show $\bar{y}$ as a function
of time together with $y$ obtained from the simulation
with $\varepsilon=0.001$. The infective increases exponentially,
and then after reaching a maximum it shows a damping
oscillation towards a nonzero value. On figure
\ref{SIRS}c we show the epidemic curve, which 
follows the same behavior with time as $y$. 
The frequency of new cases is $f=bxy$ and was obtained
from numerical simulation. Its average
$\bar{f}=b\bar{x}\bar{y}$ was also obtained from
the numerical solutions of $\bar{x}$ and $\bar{y}$.

The linearization of the equations (\ref{125b}) and (\ref{125c})
around the disease free
solution, $\bar{y}=0$, $\bar{x}=1$, and $\bar{z}=0$, gives
\beq
\frac{d\bar{y}}{dt} = \alpha\bar{y},
\eeq 
\beq
\frac{d\bar{z}}{dt} = c\bar{y} - a \bar{z}.
\eeq
The solution for $\bar{y}$ is
\beq
\bar{y} = y_0 e^{\alpha t},
\eeq
where $\alpha=b-c$. The spread occurs when $\alpha>0$,
that is, when $b>c$. Increasing the infection rate constant $b$
from small values, the threshold of the spread happens
when $b$ reaches $b_c=c$, independent of $a$.
The basic reproduction number is obtained from
(\ref{124}) and considering that $\bar{f}=b\bar{x}\bar{y}$,
and that $\alpha=b-c$, we find
\beq
R_0 = \frac{b}{c}.
\eeq

The asymptotic values $x^*$ and $y^*$ of $\bar{x}$
and $\bar{y}$ are obtained by setting to zero the
right-hand of the equations (\ref{125a}) and (\ref{125b}),
and recalling that the $\bar{z}=1-\bar{x}-\bar{y}$.
The result is
\beq
x^* = \frac{a}{b}, \qquad\qquad y^* = \frac{a(b-a)}{b(a+c)}.
\eeq
A stability analysis of this solution can also be performed.
It is possible to show that the eigenvalues related
to the stability matrix has, for some values of the
parameter, an imaginary part, which together with
a negative real part indicates a damped oscillations.
This is the behavior shown if figure \ref{SIRS} not
only for the fraction of the infective but also for
the epidemic curve.

The order parameter $s$ for the present model is
identified as the fraction $y^*$, as in the case of
the SIS model, and is given by \beq
s = \frac{a(b-c)}{c(a+c)}.
\eeq

\section{Conclusion}

We have analyzed four models of epidemic spreading using a
stochastic approach in which the primary stochastic variables
are the numbers of individuals in each class. The individuals
are classified in accordance with its condition with
respect to the infectious disease. The process of changing
from one class to the other is understood as being analogous
to a chemical reaction. This analogy allowed to use the
laws of mass action to set up the rate of several process
taking place in an epidemic spreading.

We have determined the onset of the epidemic spreading
by a linear analysis of the disease free state. From this
analysis we have determined the critical infectious rate,
above which the diseases spreads. By solving the evolution
equations, we determined the time behavior of the fraction
of the infected and the frequency of new cases. These two
quantities were also determined by numerical simulations
of the master equation.

A relevant feature of the present approach is that
the evolution equation for the average in the number of individuals
is similar to the evolution equation employed in
certain deterministic approaches. For instance, the equations
(\ref{115a}), (\ref{115b}), and (\ref{115c})
for the averages of the fractions of individuals are 
identical to those introduced by Kermack and McKendrick.
The similarity or in some cases the equality of the equations
allows to take the point of view 
according to which the stochastic approach and the
deterministic are not in opposition. Quite the contrary.
They can be understood as being consistent views
of the same problem.



\begin{thebibliography}{99}

\bibitem{bailey1957} N. T. J. Bailey, {\it The Mathematical
Theory of Epidemics}, Hafner, New York, 1957.

\bibitem{anderson1991} R. M. Anderson and R. M. May, {\it 
Infectious Diseases of Humans}, Oxford University Press,
Oxford, 1991.

\bibitem{renshaw1991} E. Renshaw, {\it Modelling Biological
Population in Space and Time}, Cambridge University Press,
Cambridge, 1991.

\bibitem{hastings1997} A. Hastings, {\it Population Dynamics},
Springer, New York, 1997.

\bibitem{keeling2008} M. J. Keeling and P. Rohani, {\it Modeling
Infectious Diseases}, Princeton University Press, Princeton, 2008.

\bibitem{bartlett1947} M. S. Bartlett,
{\it Stochastic Processes}, University of North Carolina, 1947.

\bibitem{bartlett1949} M. S. Bartlett,
J. R. Stat. Soc. B {\bf 11} 211 (1949).

\bibitem{bailey1950} N. T. Bailey, 
Biometrika {\bf 37}, 193 (1950). 

\bibitem{bailey1953} N. T. Bailey, 
Biometrika {\bf 40}, 177 (1953).

\bibitem{kermack1927} W. O. Kermack and A. G. McKendrick,
Proc. R. Soc. A {\bf 115}, 700 (1927).

\bibitem{whittle1955} P. Wittle,
Biometrika {\bf 42}, 116 (1955).

\bibitem{dietz1976} K. Dietz, in
J. Berger, W. J. Bühler, R. Repges, and P. Tautu (eds.),
{\it Mathematical Models in Medicine},
Springer, Berlin, 1976; p. 1.

\bibitem{olsen1988} L. F. Olsen, G. L. Truty, and W. M. Schaffer,
Theor. Popul. Biol. {\bf 33}, 344 (1988).

\bibitem{nisbet1982} R. M. Nisbet and W. C. S. Gurney,
{\it Modelling Fluctuating Populations}, Blackburn, Caldwell, 1982.

\bibitem{gabriel1990} J. P. Gabriel, C. Lefèvre, and P. Picard (eds.)
{\it Stochastic Processes in Epidemic Theory}
Springer, Berlin, 1990.

\bibitem{grenfell1995} B. T. Grenfell and A. P. Dobson (eds.),
{\it Ecology of Infectious Diseases in Natural Populations},
Cambridge University Press, Cambridge, 1995.

\bibitem{andersoon2000} H. Andersson and T. Britton,
{\it Stochastic Epidemic Models and their Statistical Analysis},
Springer, New York, 2000.

\bibitem{matis2000} J. H. Matis and T. R. Kiffe,
{\it Stochastic Population Models}, Springer, New York, 2000.

\bibitem{allen2015} L. J. S. Allen, {\it Stochastic Population and
Epidemic Models}, Springer, Cham, 2015.

\bibitem{britton2019} T. Britton and E. Pardoux (eds.),
{\it Stochastic Epidemic Models with Inference},
Springer, Cham, 2019.

\bibitem{tome2009} T. Tomé and M. J. de Oliveira,
Phys. Rev. E {\bf 79}, 061128 (2009).

\bibitem{tome2015L} T. Tom\'e and M. J. de Oliveira,
{\it Stochastic Dynamics and Irreversibility},
Springer, Heidelberg, 2015.

\bibitem{harris1974} T. E. Harris,
Ann. Probab. {\bf 2}, 969 (1974).

\bibitem{grassberger1983}
P. Grassberger, Math. Biosci. {\bf 62}, 157 (1983).

\bibitem{ohtsuki1986} T. Ohtsuki and T. Keyes,
Phys. Rev. A {\bf 33}, 1223 (1986).

\bibitem{satulovsky1994} J. Satulovsky and T. Tomé, 
Phys. Rev. E {\bf 49}, 5073 (1994).

\bibitem{durrett1995} R. Durrett, Spatial epidemic models,
in D. Mollison (ed.), {Epidemic Models},
Cambridge University Press, Cambridge, 1995; p. 187.

\bibitem{antal2001} T. Antal, M. Droz, A. Lipowski, and G. Odor
Phys. Rev. E {\bf 64}, 036118 (2001).

\bibitem{dammer2003} S. M. Dammer and H. Hinrichsen,
Phys. Rev. E {\bf 68}, 016114 (2003).

\bibitem{souza2010} D. R. de Souza and T. Tomé, 
Physica A {\bf 389}, 1142 (2010).

\bibitem{tome2011} T. Tomé and M. J. de Oliveira
J. Phys. A  {\bf 44}, 095005 (2011).

\bibitem{souza2013} D. R. de Souza, T. Tomé, S. T. R. Pinho,
F. R. Barreto,  and M. J. de Oliveira, 
Physical Review E 87, 012709 (2013).

\bibitem{tome2015} A. H. O. Wada, T. Tomé, and M. J. de Oliveira,
J. Stat. Mech. P04014 (2015).

\bibitem{ruziska2017} F. M. Ruziska, T. Tomé, M. J. de Oliveira,
Physica {\bf 467}, 21 (2017).

\bibitem{kampen1981} N. G. van Kampen, {\it Stochastic Processes
in Physics and Chemistry}, North-Holland, Amsterdam, 1981.

\bibitem{kampen1961} N. G. van Kampen, Can. J. Phys. {\bf 39}, 551 (1961).

\bibitem{kampen1973} N. G. van Kampen, Biometrika {\bf 60}, 419 (1973). 

\bibitem{mcneil1972} D. R. McNeil, Biometrika {\bf 59}, 494 (1972).

\bibitem{ross1911} R. Ross, {\it The Prevention of Malaria},
Murray, London, 1911.

\bibitem{heesterbeek2015} J. A. P. Heesterbeek and M. G. Roberts,
Phil. Trans. R. Soc. B {\bf 370}, 20140307 (2015).

\end{thebibliography}
\end{document}